\documentstyle[11pt,epsf]{article}
\begin{document}

\hspace*{-1.9em} \parbox[t]{4.7in}{ {\Large {\bf Noise properties of the 
SET transistor in the co-tunneling regime} } } 

\vspace*{3ex}

D.V. Averin 

\vspace*{1ex}

\parbox[t]{4in}{ {\em Department of Physics and Astronomy, SUNY, 
Stony Brook \\ NY 11794-3800} }

\vspace*{2ex}

{\bf Abstract.} Zero-frequency spectral densities of current 
noise, charge noise, and their cross-correlation are calculated 
for the SET transistor in the co-tunneling regime. The current 
noise has a form expected for the uncorrelated co-tunneling 
events. Charge noise is created by the co-tunneling and also by 
the second-order transitions in a single junction. Calculated 
spectral densities determine transistor characteristics as 
quantum detector. 

\vspace*{6ex}

\hspace*{-1.9em} {\bf 1. INTRODUCTION}   

\vspace*{1ex}

Single-electron-tunneling (SET) transistor [1,2,3] is the 
natural measuring device for the potential quantum logic circuits 
based on the charge states of mesoscopic Josephson junctions [4,5,6]. 
This fact, together with the general interest to the problem of 
quantum measurement, motivates current discussions of the SET 
transistors as quantum detectors -- see, e.g., [7]. Detector 
characteristics of the transistor are determined by its 
noise properties, and have been studied so far [8,9] in the 
regime of classical electron tunneling. The aim of this brief 
note is to calculate the transistor noise properties in the 
Coulomb blockade regime, when the current flows in it
by the process of co-tunneling. An expected, and confirmed 
below, advantage of co-tunneling for quantum signal 
detection is a weaker back-action noise on the measured system 
produced by the SET transistor.

\vspace*{4ex}

\hspace*{-1.9em} {\bf 2. CO-TUNNELING IN THE SET TRANSISTOR} 

\vspace*{1ex}

SET transistor [1] is a small conductor, typically a small 
metallic island, placed between two bulk external electrodes, 
that forms two tunnel junctions with these electrodes. Due to 
Coulomb charging of the island by tunneling electrons, the 
current $I$ through the structure depends on the island 
electrostatic potential controlled by an external gate voltage 
$V_g$. Sensitivity of the current $I$ to variations of the 
voltage $V_g$ makes it possible to measure this voltage, 
and is the basis for transistor operation as the detector. 

When the bias voltage $V$ across the transistor is smaller than 
the Coulomb blockade threshold (dependent on $V_g$), the 
tunneling is suppressed by the Coulomb charging energy required 
to transfer an electron in or out of the central electrode. In 
this regime, the current $I$ flows through the transistor only 
by the co-tunneling process that consists of two electron jumps 
across the two transistor junctions in the same direction (for 
review, see [10]). Energy diagram of the co-tunneling 
transitions is shown in Fig.\ 1. Transitions go via two virtual 
intermediate charge states with $n=\pm 1$ extra electrons on 
the island and large charging energies $E_{1,2}$. The energies 
$E_{1,2}$ are equal to the change of electrostatic energy due 
to the first electron jump in the first or the second junction, 
and depend on the voltages $V$, $V_g$, junction capacitances 
$C_{1,2}$, and capacitance $C_g$ that couples the gate voltage 
$V_g$: 
\begin{equation} 
E_1 = E_C(\frac{1}{2}+q_0-(C_1+\frac{C_g}{2})\frac{V}{e})\, , 
\;\;\;\;  E_2 = E_C(\frac{1}{2}-q_0-(C_2+\frac{C_g}{2}) 
\frac{V}{e}) \, . 
\label{1} \end{equation} 
Here $E_C$ is the characteristic charging energy of the transistor 
$E_C =e^2/C_{\Sigma}$, $C_{\Sigma}\equiv C_1+C_2+C_g$, and $q_0$ is 
the charge (in units of electron charge $e$) induced by the gate 
voltage, $q_0\equiv C_gV_g/e$. Equations (\ref{1}) assume
that the voltage $V_g$ is applied to the transistor symmetrically.  

\begin{figure}[htb]
\setlength{\unitlength}{1.0in}
\begin{picture}(5.,1.0) 
\put(.3,.0){\epsfxsize=4.6in\epsfysize=0.8in\epsfbox{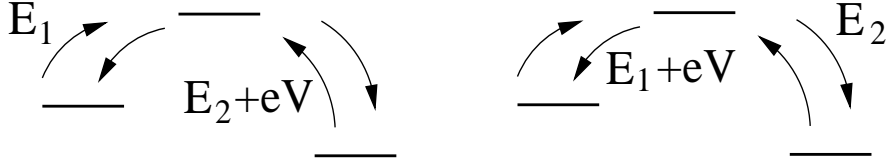}}
\end{picture}
\caption{Energy diagrams of the four co-tunneling transitions in 
the SET transistor. The arrows indicate electron jumps. The left 
and right diagrams show, respectively, the transitions through 
the charge states with $n=+1$ and $n=-1$ extra electrons on the 
central electrode of the transistor. The energy labels are the 
changes of electrostatic energy in the first electron jump of 
each co-tunneling process. }
\end{figure} 

The co-tunneling regime is realized for values of $V$ and $V_g$ 
that make the energies $E_j, \,j=1,2$, sufficiently large, $E_j\gg T, 
\hbar G_{1,2}/C_{\Sigma}$, where $T$ is the temperature and $G_{1,2}$ 
are the tunnel conductances of the transistor junctions. The 
conductances $G_{1,2}$ are assumed to be small, $g_{1,2} \equiv 
\hbar G_{1,2}/2\pi e^2 \ll 1$. In this regime, one can 
neglect thermally-induced ``classical'' electron transitions of the 
first order in conductances $g_j$, and also neglect quantum 
broadening of the charge states by tunneling [11,12]. Transport 
characteristics of the transistor are determined then by the 
co-tunneling transitions of the second order in $g_j$.  

The gate voltage $V_g$ controls the co-tunneling current $I$ in 
the transistor through the dependence of the energies $E_j$ on the 
induced charge $q_0$. Viewed as the detector for measurement of small 
variations of $V_g$, the transistor is characterized by: (1) the 
linear response coefficient $\lambda = \delta \langle I\rangle/ 
\delta V_g$, where $\delta\langle I\rangle$ is the change of the 
average current due to variation of $V_g$; (2) the 
spectral density $S_I$ of the current noise (output noise of the 
detector); (3) the spectral density $S_Q$ of the fluctuations of 
the charge $Q=en$ on the central electrode of the transistor 
(``back-action'' noise); and (4) cross-correlation $S_{IQ}$ between 
the current and charge noise. In general, the back-action noise 
is the fundamental property of a quantum detector 
responsible for localization of the measured system in the 
eigenstates of the measured observable. For the SET transistor, the 
measured observable is the voltage $V_g$, and the back-action 
noise originates from fluctuations of the charge $Q$ in the process 
of electron transfer through the transistor. The voltage $V_g$ 
is coupled to $Q$ in the transistor energy as $QV_gC_g/C_{\Sigma}$, 
where $Q=0, \pm e$ in the co-tunneling regime. The fluctuations of 
$Q$ produce random force that acts on the system creating $V_g$, 
and lead to mutual decoherence of the states of this system with 
different values of $V_g$. 

In the next section, the noise spectral densities are calculated 
in the zero-frequency limit for the SET transistor in the 
co-tunneling regime. The zero-frequency results determine the 
detector characteristics of the transistor in the frequency range 
below the characteristic frequency of electron tunneling. 

\vspace*{4ex}

\hspace*{-1.9em} {\bf 3. NOISE CALCULATION} 

\vspace*{1ex}

Transport properties of the transistor related to co-tunneling 
can be calculated by straightforward perturbation theory in tunneling 
in the second order in tunnel conductances $g_j$. The tunneling part 
of the transistor Hamiltonian is 
\[ H_T=\sum_{j=1,2} H^{\pm}_j \, , \] 
where $H^{\pm}_j$ describe forward and backward electron transfer    
in the $j$th junction. The junction electrodes are assumed to 
have quasicontinuous density of states. The only nonvanishing 
free correlators of $H^{\pm}_j$ can be expressed in terms of the 
conductances $g_j$ (see, e.g., [10]): 
\begin{equation} 
\langle H_j^{\pm}(t) H^{\mp}_j(t') \rangle = g_j 
\int \frac{d\omega \omega e^{i\omega (t'-t)} }{1-e^{-\omega /T}}  
\, .  
\label{2} \end{equation} 

The zero-frequency spectral densities of current and charge noise 
are given by the standard expressions. For instance, 
\begin{equation} 
S_I = \frac{1}{\pi} \mbox{Re} \int ^{t}_{-\infty} dt' 
\mbox{Tr} \{ S^{\dagger}(t) I(t) S(t,t') I(t') S(t') \rho_0 \}\, , 
\label{3} \end{equation}
where $S(t,t')={\cal T} \exp \{ -i\int^{t}_{t'} d\tau H_T(\tau) 
\}$ is the evolution operator of the transistor due to tunneling, 
$\rho_0$ is its unperturbed equilibrium density matrix, $S(t) 
\equiv S(t,-\infty)$, and the current $I$ can be calculated 
in either of the two junctions, e.g., $I=ie(H^+_1-H^-_1)$. The 
time dependence of all operators $H_T$ is now due to both 
internal energies of the junction electrodes and the electrostatic 
charging energy of the transistor. The term $-\langle I 
\rangle^2$ is omitted in (\ref{3}), since the average co-tunneling 
current $\langle I\rangle$ is of the same order in $g_j$ as 
$S_I$. Therefore, $\langle I\rangle^2$ is of higher order than 
$S_I$, and can be neglected in the perturbative calculation 
adequate for the co-tunneling regime. The same considerations apply 
to the spectral densities $S_Q$ and $S_{IQ}$ discussed below. 

Expanding all evolution operators in (\ref{3}) up to the second 
order in $H_j^{\pm}$, and evaluating the averages with the help of 
Eq.\ (\ref{2}), we find that $S_I$ is given by the standard 
expression characteristic for uncorrelated tunneling: 
\begin{equation}  
S_I= \frac{e^2}{2\pi} (\gamma^+ +\gamma^- ) \, . 
\label{4} \end{equation}
Here $\gamma^{\pm}$ are the rates of forward and backward 
co-tunneling:
\[ \gamma^+ = \frac{2 \pi g_1g_2}{\hbar } \int \frac{d\omega_1 
\omega_1}{1- e^{- \omega_1 /T} } \frac{d\omega_2 \omega_2}{1- 
e^{- \omega_2 /T} } \delta (\omega_1+\omega_2-eV) 
\left( \frac{1}{E_1+\omega_1}+\frac{1}{E_2+\omega_2} \right)^2 
\, , \] 
and $\gamma^-$ is given by the same expression with $eV$ and $E_j$ 
changed into $-eV$ and $E_j+eV$. The rates $\gamma^{\pm}$ also 
determine the average co-tunneling current, $\langle I\rangle 
= e(\gamma^+ -\gamma^- )$. 

Expression for the spectral density $S_Q$ of charge fluctuations is 
obtained from Eq.\ (\ref{3}) by replacing the current operators with 
the charge operators $Q=en$. To find $S_Q$ in the co-tunneling regime 
one needs to expand the evolution operators in this expression up to 
the fourth order in the tunneling terms $H_T$. The non-vanishing 
contribution to $S_Q$ comes then from one particular choice of orders 
of expansion of different evolution operators: 
\begin{eqnarray} 
S_Q= - \frac{e^2}{\pi} \mbox{Re} \int ^{t}_{-\infty} dt_1 
\int ^{t}_{-\infty} dt_2\int ^{t_2} dt_3 \int ^{t_3} dt'\int ^{t'} 
dt_4 \nonumber \\  
\mbox{Tr} \{ H_T(t_1) n H_T(t_2) H_T(t_3) n H_T(t_4) \rho_0 \}\, .  
\label{5} \end{eqnarray}
Taking the trace in Eq.\ (\ref{5}) we get the two types of 
contributions to $S_Q$. In one of them, the pairings of operators 
$H_T$ belong to the two different junctions, while in the other, 
all $H_T$'s belong to one junction. The first contribution describes 
fluctuations of the charge in the process of co-tunneling, and can 
be split into the two terms, $S^{\pm}$, associated with the forward 
and backward transitions. The second contribution $\bar{S}$ 
is created by the back-and-forth tunneling within the same 
junction. Accordingly, $S_Q$ can be written as   
\begin{equation}  
S_Q= \bar{S}+S^{\pm} \, , 
\label{6} \end{equation} 
where the different terms are found from Eq.\ (\ref{5}) to be
\[ \bar{S} = \sum_j (2\pi eg_j)^2 
\frac{\hbar T^3}{6}  \left(\frac{1}{E_j^2}-\frac{1}{(E_{j'}+eV)^2} 
\right)^2 \, , \;\;\;\;\; j,j'=1,2\, , \;\; j'\neq j \, , \] 
\[ S^+ = \hbar e^2 g_1g_2\int \frac{d\omega_1 
\omega_1}{1- e^{- \omega_1 /T} } \frac{d\omega_2 \omega_2}{1- 
e^{- \omega_2 /T} }\delta (\omega_1+\omega_2-eV) \cdot \] 
\[ \left( \frac{1}{(E_1+\omega_1)^2} -\frac{1}{(E_2+\omega_2)^2} 
\right)^2 \, . \] 
The last term $S^-$ is given by the same expression as $S^+$ with 
$eV$ and $E_{1,2}$ replaced, respectively, by $-eV$ and 
$E_{1,2}+eV$. 

Finally, we calculate the correlator $S_{IQ}$ between the charge 
and current noise:  
\begin{equation} 
S_{IQ} = \frac{1}{2 \pi} \int dt' \mbox{Tr} 
\{ S^{\dagger}(t) I(t) S(t,t') Q(t') S(t') \rho_0 \}\, . 
\label{7} \end{equation}
Similarly to Eq.\ (\ref{3}), it is convenient to break the integral 
in (\ref{7}) into the two parts, $t'<t$ and $t'>t$. Expanding 
then the evolution operators up to the third order in $H_T$, and 
evaluating averages using Eq.\ (\ref{2}), we find: 
\begin{eqnarray} 
S_{IQ} = i e^2 g_1g_2 \int \frac{ d\omega_1 
\omega_1}{1- e^{- \omega_1 /T} } \frac{d\omega_2 \omega_2}{1- 
e^{- \omega_2 /T} } \biggl\{ \delta (\omega_1+\omega_2-eV) \cdot 
\nonumber \\
\left( \frac{1}{E_1+\omega_1} - \frac{1}{E_2+\omega_2} \right) 
\left( \frac{1}{E_1+\omega_1} +\frac{1}{E_2+\omega_2} \right)^2 + 
\delta (\omega_1+\omega_2+eV) \cdot \label{8} \\
\left( \frac{1}{E_2+eV+\omega_2} - \frac{1}{E_1+eV+\omega_1} 
\right) \left( \frac{1}{E_1+eV+ \omega_1} +\frac{1}{E_2+eV+\omega_2} 
\right)^2 \biggr\} \, . \nonumber 
\end{eqnarray}

The correlator $S_{IQ}$ (\ref{8}) is purely imaginary. From the 
perspective of the general theory of quantum linear detection [13] 
this fact means that the SET transistor in the co-tunneling regime 
is ``symmetric'' detector with the output noise and back-action 
noise uncorrelated at the classical level, since classically, the 
correlator between current and charge is given by the symmetrized 
correlation function which contains only the real part of $S_{IQ}$. 
Imaginary part of $S_{IQ}$ should be directly related to the linear 
response coefficient $\lambda$ of the transistor [13]. Comparison 
of Eq.\ (\ref{8}) to the expression for $\lambda = \delta \langle 
I\rangle/ \delta V_g$ shows that, indeed, the two quantities are  
related: $(C_g/C_{\Sigma}) \mbox{Im}[S_{IQ}]= -\hbar \lambda/4\pi$. 

\vspace*{4ex}

\hspace*{-1.9em} {\bf 4. RESULTS AND DISCUSSION} 

\vspace*{1ex}

In this section, we calculate explicitly the noise spectral 
densities $S_Q$, $S_I$, and $S_{IQ}$ in various limits, and discuss 
the implications of the obtained relations for the characteristics 
of transistor as quantum detector. At small bias voltages $V$, 
$V\sim T \ll E_j$, the electron excitation energies 
$\omega_j$ can be neglected in comparison to energy barriers $E_j$, 
and expressions for the spectral densities obtained in the previous 
section are simplified: 

\[ S_Q= \frac{\hbar e^2}{6} \left( \frac{1}{E_1^2}- \frac{1}{E_2^2} 
\right)^2 \Bigl\{ g_1g_2[(eV)^2+(2\pi T)^2] eV \coth \frac{eV}{2T} 
+(g_1^2 +g_2^2)4\pi^2 T^3 \Bigr\} \, , \]

\vspace{-2em}

\begin{eqnarray}  
S_I= \frac{g_1g_2e^3V}{6 \hbar} \left( \frac{1}{E_1}+ 
\frac{1}{E_2} \right)^2 [(eV)^2+(2\pi T)^2] \coth \frac{eV}{2T} \, , 
\label{10}  \\
S_{IQ} = i \frac{g_1g_2e^3V}{6 } \left( \frac{1}{E_1}- 
\frac{1}{E_2} \right)\left( \frac{1}{E_1}+\frac{1}{E_2} \right)^2 
[(eV)^2+(2\pi T)^2] \, . \nonumber
\end{eqnarray} 

A fundamental figure-of-merit of a quantum detector is ``energy 
sensitivity'' $\epsilon$ that is defined in terms of the output 
noise, back-action noise, and response coefficient of the 
detector - see, e.g., [13]. Qualitatively, energy sensitivity 
characterizes the amount of noise introduced by the detector into 
the measurement process, and is limited from below by $\hbar/2$ 
due to the quantum mechanical restrictions on the detector 
dynamics. In the case of SET transistor in the co-tunneling regime 
(when the current $I$ and charge $Q$ are classically uncorrelated), 
$\epsilon$ can be expressed directly through the three spectral 
densities $S_I$ (\ref{4}), $S_Q$ (\ref{6}), and $S_{IQ}$ (\ref{8}),  
as follows:  
\begin{equation} 
\epsilon=\frac{\hbar}{2} \frac{\sqrt{S_IS_Q}}{|S_{IQ}|} 
\, , 
\label{11} \end{equation}
Equations (\ref{10}) show that at small bias voltages the 
transistor energy sensitivity (\ref{11}) is: 
\begin{equation} 
\epsilon = \frac{\hbar}{2} \left[ \coth \frac{eV}{2T} \Biggl( \coth 
\frac{eV}{2T} + \biggl(\frac{g_1}{g_2}+\frac{g_2}{g_1} \biggr) 
\frac{4\pi^2 T^3}{ eV[(eV)^2+(2\pi T)^2] } \Biggr) \right]^{1/2}
 \, .  
\label{12} \end{equation}
The energy sensitivity (\ref{12}) approaches the fundamental 
limit $\hbar/2$ at $eV\gg T$ [14]. 

At larger bias voltages $eV\sim E_j \gg T$, temperature $T$ can be 
neglected in equations for the noise spectral densities of the 
previous section and they give: 
\begin{eqnarray} 
S_Q=\hbar e^2 g_1g_2 \Biggl\{ \frac{(eV)^3}{6} \sum_{j} 
\frac{1}{(E_j(E_j+eV))^2} +\frac{4eV}{(E_1+E_2+eV)^2} - 
\nonumber \\ 
2\frac{eV(E_1+E_2)+(eV)^2+2E_1E_2}{(E_1+E_2+eV)^3} \sum_{j} 
\ln \biggl( 1+\frac{eV}{E_j} \biggr) \Biggr\}  \, , 
\nonumber \\
S_I=\frac{e^2 g_1g_2}{\hbar } \Biggl\{ \biggl(eV+\frac{2E_1E_2 
}{E_1+E_2+eV} \biggr) \sum_{j} \ln \biggl( 1+\frac{eV}{E_j} \biggr) 
- 2eV\Biggr\} \, , \label{13} \\
S_{IQ} = ie^2 g_1g_2 \Biggl\{ \frac{eV(E_1-E_2) }{(E_1+eV)(E_2+eV)} 
\biggl(1+\frac{eV(E_1+E_2+eV)}{2E_1E_2} \biggr) - \nonumber \\ 
\frac{E_1-E_2}{E_1+E_2+eV}
\sum_{j} \ln \biggl( 1+\frac{eV}{E_j} \biggr) \Biggr\}  \, .
\nonumber \end{eqnarray} 
Figure 2 shows the zero-temperature energy sensitivity (\ref{11}) 
as a function of the bias voltage calculated from these equations 
for the SET transistor with equal junction capacitances and several 
values of the charge $q_0$ induced by the gate voltage. The energy 
sensitivity reaches $\hbar/2$ at small bias voltages and diverges 
when the voltage approaches the Coulomb blockade threshold (see 
also Eq.\ (\ref{14}) below). The non-monotonic behavior of $\epsilon$ 
at small $q_0$ is caused by the fact that the correlator $S_{IQ}$ 
vanishes for $E_1\rightarrow E_2$. 

\begin{figure}[htb]
\setlength{\unitlength}{1.0in}
\begin{picture}(3.4,2.3) 
\put(0.9,.0){\epsfxsize=3.4in\epsfysize=2.2in\epsfbox{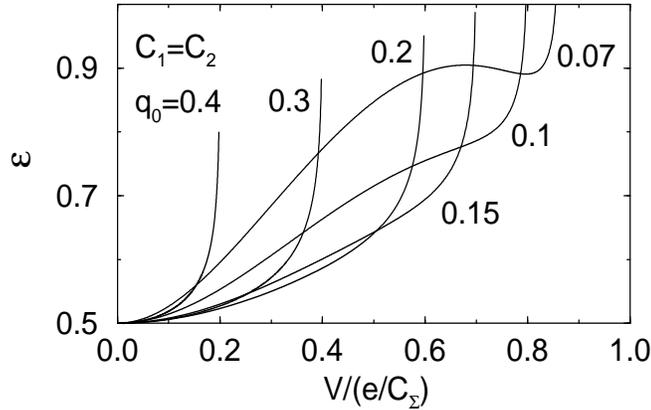}}
\end{picture}
\caption{Energy sensitivity, in units of $\hbar$, of the SET 
transistor in the co-tunneling regime. }
\end{figure}

The energy barriers $E_j$ decrease with increasing bias voltage $V$,   
and one or both of them disappear when $V$ approaches the Coulomb 
blockade threshold. At the threshold, the spectral densities (\ref{13}) 
diverge. For generic values of the gate voltage, only one of the 
barriers is suppressed at the threshold, e.g., $E_1\rightarrow 0$. 
In this case, 
\[ S_Q= \frac{\hbar g_1g_2 e^3V}{6E_1^2} \, , \;\;\; 
S_I= \frac{g_1g_2 e^3V}{\hbar} \ln \left( \frac{eV}{E_1} \right) 
\, , \;\;\; S_{IQ}= -i \frac{g_1g_2 e^3V}{2E_1 } \, . \] 
The energy sensitivity (\ref{11}) also slowly diverges:   
\begin{equation} 
\epsilon = \frac{\hbar}{2} \left[ \frac{2}{3} \ln \left( 
\frac{eV}{E_1} \right)\right]^{1/2}  \, .  
\label{14} \end{equation}
All these divergencies should be regularized by the finite 
width of the charge states created by tunneling [11,12]. 
Qualitatively, this means that for $E_1$ smaller than  $\hbar 
G_2/C_{\Sigma}$, the logarithm in Eq.\ (\ref{14}) should 
saturate at $\ln (1/g_2)$. Quantitative treatment of the 
threshold region is outside the scope of this work.  

To summarize, we have studied noise properties of the SET transistor 
in the co-tunneling regime, and calculated its energy sensitivity 
as a detector. The energy sensitivity approaches $\hbar/2$ for 
small bias voltages, and slowly diverges at the Coulomb blockade 
threshold. 

\vspace*{.4ex} 

This work was supported by AFOSR. 

\vspace*{3ex}

\hspace*{-.3em}{\bf REFERENCES}   

\renewcommand{\labelenumi}{\hspace*{-1.9em}[\theenumi]}

\begin{enumerate}

\item D.V. Averin and K.K. Likharev, J.\ Low Temp.\ Phys. {\bf 62}, 
345 (1986). 

\item T.A. Fulton and G.J. Dolan, Phys.\ Rev.\ Lett {\bf 59}, 109 
(1987).

\item M.A. Kastner, Rev.\ Mod.\ Phys. {\bf 64}, 849 (1992).

\item D.V. Averin, {\it Solid State Commun.} {\bf 105}, 659 (1998).

\item Yu. Makhlin, G. Sch\"{o}n, and A. Shnirman, Nature {\bf 398}, 
305 (1999). 

\item Y. Nakamura, Yu.A. Pashkin, and J.S. Tsai, Nature {\bf 398}, 
786 (1999). 

\item M.H. Devoret and R.J. Schoelkopf, Nature {\bf 406}, 1039 
(2000).
 
\item A. Shnirman and G. Sch\"{o}n, Phys.\ Rev. B {\bf 57}, 15400 
(1998). 

\item A.N. Korotkov, cond-mat/0008461.

\item D.V. Averin and Yu.V. Nazarov, in: {\em ``Single Charge 
Tunneling''}, Ed.\ by H. Grabert and M. Devoret (Plenum, NY, 1992), 
p.\ 217.  

\item A.N. Korotkov, D.V. Averin, K.K. Likharev, and S.A. Vasenko, 
in: {\em ``Single Electron Tunneling and Mesoscopic devices''}, Ed. 
by H. Koch and H. L\"ubbig, (Springer, Berlin, 1992), p.\ 45.

\item Yu.V. Nazarov, J.\ Low Temp.\ Phys.\ {\bf 90}, 77 (1993).

\item D.V. Averin, in: {\em ``Exploring the Quantum-Classical 
Frontier: Recent Advances in Macroscopic and Mesoscopic Quantum 
Phenomena''}, Eds. J.R. Friedman and S. Han, to be published; 
cond-mat/0004364. 

\item It should be noted, however, that this fact does not 
necessarily mean that the regime of relatively small bias voltages, 
$T\ll eV \ll E_j$, represents optimal operating point of the 
practical SET transistors. Other noise sources, not included 
in the model but present in realistic systems, make it important 
to have large absolute values of the output signal, the condition 
that is not fulfilled for the SET transistor biased deep inside the 
Coulomb blockade region.  

\end{enumerate}
\end{document}